\documentclass[9pt,twocolumn,twoside]{opticajnl}
\journal{opticajournal} 

\setboolean{shortarticle}{true}

\usepackage[skip=25pt]{caption} 

\usepackage{lineno} 

\title{Dramatic Enhancement of Supercontinuum Generation in H$_2$O by Non-Harmonic Two-Color Excitation
} 

\author[1,*]{Tsuneto Kanai}
\author[1,2]{ChengXiang Jin}
\author[1,2]{Hibiki Tsunekawa}
\author[1,2,3]{Atsunori Sakurai}
\author[1,2,3]{Toshiki Sugimoto}
 
\affil[1]{Institute for Molecular Science, National Institutes of Natural Sciences,
Okazaki, Aichi 444-8585, Japan}
\affil[2]{Graduate Institute for Advanced
Studies, SOKENDAI, Okazaki, Aichi 444-8585, Japan}
\affil[3]{Laser-Driven Electron-Acceleration Technology Group,
RIKEN SPring-8 Center, Sayocho, Hyogo 679-5148,
Japan}
 
\affil[*]{goldwell74@gmail.com}

\begin{abstract}
We demonstrate a dramatic enhancement of supercontinuum generation (SCG) in H$_2$O driven 
by non-harmonic two-color laser fields, where pump and seed frequencies are not integer multiples. 
This scheme yields a broadband SC spectrum enhanced by approximately three orders of magnitude 
over the single-color case. 
Systematic experiments and theoretical analysis, including isotope effects, 
show that the enhancement arises from soliton compression with resonant dispersive wave (RDW) emission, 
phase- and group-velocity-matched four-wave mixing (FWM), and cross-phase modulation (XPM), 
offering a versatile route to stronger SCG and noise suppression in broadband spectroscopy in H$_2$O,
an emerging topic in attosecond  and surface physics.
\end{abstract}

\setboolean{displaycopyright}{false} 

\begin{document}
 
\maketitle

SCG from intense ultrashort pulses in transparent media is a key light source  in ultrafast optical science 
\cite{Alfano1970a,DAWES1969,Dubietis2019,Chin2010,Liu2005,Marburger1975},
supporting applications including seed sources for optical parametric amplifiers (OPAs) \cite{Kanai2019,Kanai2017}, time-resolved spectroscopy \cite{Hassan2016}, frequency combs \cite{Udem2002}, nonlinear microscopy \cite{Paulsen2003}, and lightwave engineering \cite{Goulielmakis2007}. 
Recent high-power mid-infrared (mid-IR) laser systems, especially those using SCG from picosecond pulses as a seed \cite{Kanai2019,Kanai2017}, 
have expanded 
the scope of SCG beyond conventional $\chi^{(3)}$-based processes 
into  high-order and strongly nonperturbative phenomena such as high harmonic generation (HHG) \cite{Zheltikov2019,Popmintchev2012,Takahashi2010} 
and high-order frequency conversion \cite{Nomura2007}. 
The interplay between HHG and SCG has attracted growing interest.

Harmonics-enhanced SCG  \cite{Bejot2014,Garejev2016,Loures2015,Vengris2019,Mitrofanov2015,Akoezbek2002}, a representative example of this interplay, 
has been extensively explored using two-color filamentation schemes with a second harmonic (SH) \cite{Vengris2019,Nomura2015,Cook2000,Zheng2008,Zhao2015}. 
When the SH field efficiently breaks inversion symmetry \cite{Kanai2001}, 
this configuration can enable efficient SCG from the THz to soft-x-ray region \cite{Cook2000,Zhao2015,Nomura2015, Zheng2008}.
In contrast, few studies have explored two-color fields with non-integer frequency ratios, 
termed \textit{non-harmonic two-color fields} here~\cite{Theberge2006a,Junaid2024,Liu2013,Ensley2011,Wang2006,Takahashi2010,Kanai2017}.
Prior work demonstrated tunable visible pulses via FWM in gases~\cite{Theberge2006a} and broadband sidebands via cascaded FWM in BBO~\cite{Liu2013}, 
but systematic studies in liquids are scarce: Most recently, Junaid \textit{et al.} investigated single and cascaded FWM 
as well as stimulated Raman scattering (SRS) in liquid-core fibers using continuous-wave and picosecond laser sources \cite{Junaid2024}.

In this Letter, we demonstrate a dramatic enhancement (DE) of SCG in H$_2$O using non-harmonic two-color laser fields. 
Compared to single-color or harmonic excitation, this configuration can yield much stronger SCG. 
Systematic experiments and theoretical analysis, including isotope effects, 
reveal that the enhancement originates from a synergistic interplay among soliton compression 
accompanied by RDW emission \cite{Leo2014,Genty2004,Vasa2014,Suminas2016,Panagiotopoulos2015,Heinzerling2025}, 
phase- and group-velocity-matched FWM \cite{Dharmadhikari2016}, and subsequent cascaded FWM and XPM. 
These findings not only introduce a versatile approach to achieving
broader and more intense SCG, but also provide an effective means of suppressing the background noise signal, 
an inherent issue in broadband/multicolor spectroscopy in liquids--particularly in H$_2$O, 
an emerging topic in attosecond science~\cite{Pupeza2020,Heinzerling2025} and surface/interface physics~\cite{Sugimoto2016,Kanai2025CleoGra}.

\begin{figure}[h]
\centering
\includegraphics[width=0.9\linewidth]{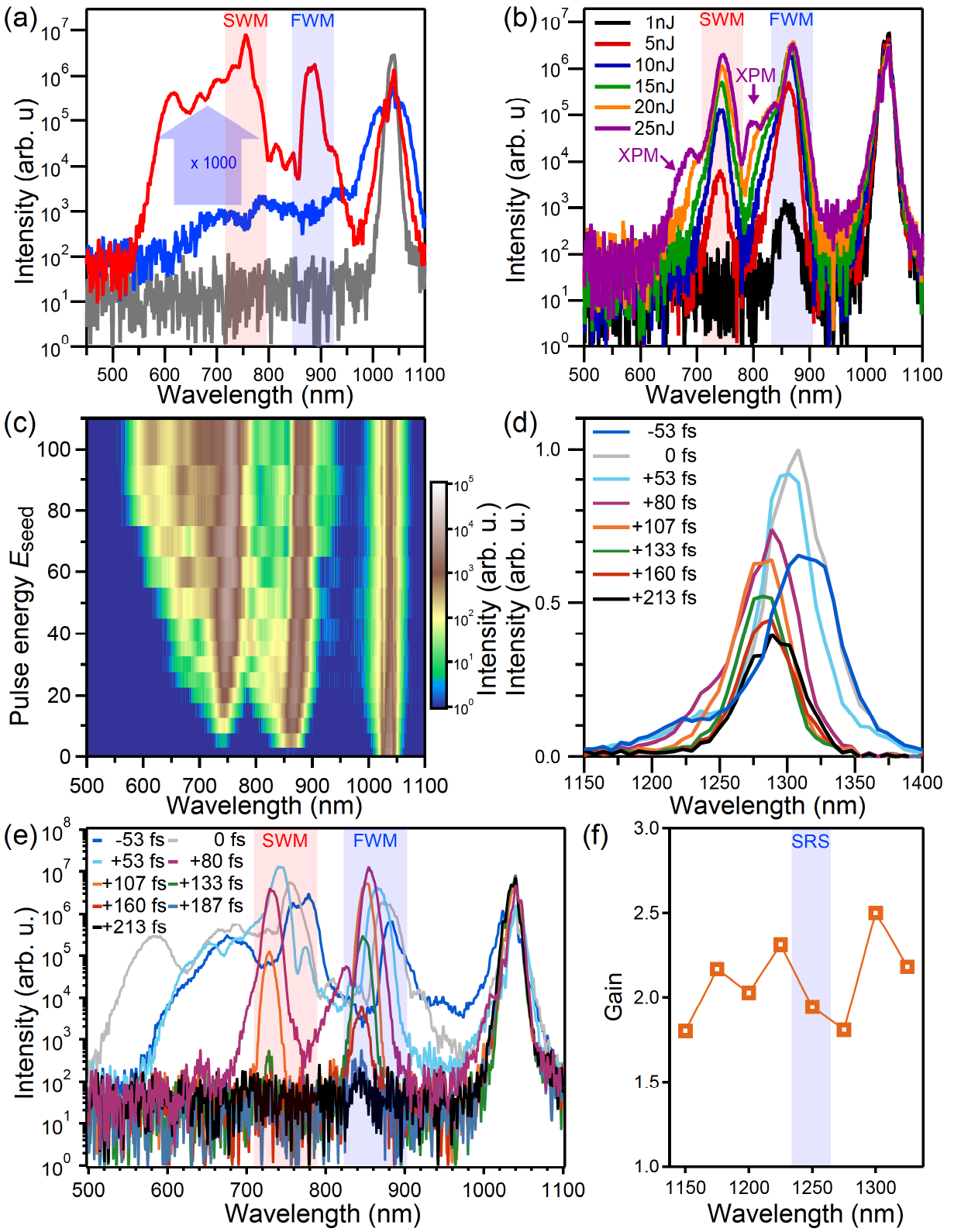}
\caption{
\small{ 
(a) SC spectra generated in H$_2$O  (red), showing strong enhancement when driven by both the pump ($1036$~nm, 800~nJ) and seed ($1300$~nm, 100~nJ) pulses. 
For comparison, single-color SC spectra are shown for the same pump energy (800~nJ, gray) and for an increased pump energy (1200~nJ, blue).  
(b, c) Dependence of the SC spectra on the seed pulse energy $E_\mathrm{seed}$ in the low-energy regime [1--25~nJ, (b)] and over an extended range up to 100~nJ (c).  
(d, e) SC spectra measured using a Si-based spectrometer (d), and amplified seed spectra measured using an InGaAs-based spectrometer (e), 
in H$_2$O at various time delays $\tau$ (from $-53$~fs to $+213$~fs). 
Distinct spectral peaks attributed to SWM and FWM are observed in the SWM and FWM regions, respectively.
(f) Spectral gain of the seed pulse, defined as the output-to-input intensity ratio, 
plotted as a function of $\lambda_\mathrm{seed}$ in H$_2$O.
}
}
\label{fig1}
\end{figure}

The experiments employed a Yb:KGW chirped-pulse amplification laser system (CARBIDE-CB3, Light Conversion Ltd.)
delivering 180-fs, 1036-nm pulses at up to 400~{\textmu}J and 2~MHz, while we operated the system at 40~{\textmu}J and 100~kHz for this study.
The output beam was split into two parts: one part, equipped with a delay line, served as the pump for the two-color SCG experiments, 
while the other, driving an OPA (ORPHEUS-F, \textit{ibid}), was used to generate the seed for these experiments, 
producing 640--940~nm and 1.15--2.5~{\textmu}m pulses with durations of 40--80~fs and energies of 0.4--1.6~{\textmu}J.
The pump and seed beams were spatially overlapped using dichroic mirrors and focused collinearly 
into the center of a 7.4-mm-thick layer of liquid H$_2$O or its isotopologue D$_2$O, sandwiched between 1.3-mm-thick fused silica windows, 
by a lens with a focal length of 150~mm. 
Neutral density filters were used to control the pulse energy of each beam, and telescopic lens systems were inserted into both beam paths 
to ensure nearly identical focal spot sizes, with a $1/e^2$ radius $w_0$ of approximately 34~{\textmu}m 
in  H$_2$O and D$_2$O.
The peak power of the pump was typically set to 4.2~MW by adjusting the pulse energy to 800 nJ, 
$\approx 67\%$ of the critical power for filamentation,
$P_\mathrm{cr} = {3.77 \lambda^2}/(8\pi n_0 n_2)$~(theoretically predicted by Marburger \cite{Marburger1975} 
and experimentally demonstrated by Liu and Chin \cite{Liu2005}),
which amounts to 6.3~MW for H$_2$O assuming $n_2 = 1.9 \times 10^{-16}$~cm$^2$/W~\cite{Tcypkin2021}.
The generated SC spectra were recorded by collecting Fresnel reflections from a fused silica prism 
with InGaAs-based (NIRQuest-256, Ocean Optics Inc.; 900--2500 nm) and Si-based (LR1, Aseq Instruments; 200--1100 nm) spectrometers.
 

We consider generalized FWM between an intense pump and a weak seed, along with its cascaded processes. 
When the seed frequency is an integer multiple of the pump 
($\omega_\mathrm{seed} = q\,\omega_\mathrm{pump}$, $q \in \mathbb{Z},\ q \geq 2$), 
as in harmonics-enhanced SCG, the FWM-generated waves 
($\omega_\mathrm{FWM} = |q \pm 2|\,\omega_\mathrm{pump},\ |2q \pm 1|\,\omega_\mathrm{pump}$) 
retain harmonic relationships with the pump.
Here, four-wave rectification \cite{Cook2000,Zhao2015,Nomura2015} 
is regarded as the case  ``$\omega_\mathrm{FWM} = 0$'' with $q = 2$.
Consequently, higher-order processes dominantly realized through its cascading, such as effective six-wave mixing (SWM) 
and eight-wave mixing (EWM), also generate only harmonically related frequencies.
In contrast, for non-harmonic pump-seed pairs, the generated frequencies become intrinsically non-harmonic, 
dramatically increasing the number of accessible frequency components. 
This enhanced effect is analogous to the generation of dissonance by two non-harmonic sounds.
When one sets $\omega_\mathrm{pump}/2 < \omega_\mathrm{seed} < \omega_\mathrm{pump}$,
degenerate FWM generates an idler--not to be confused with the idler of OPAs--on the anti-Stokes side at 
$\omega_\mathrm{FWM}^\mathrm{as} = 2\omega_\mathrm{pump} - \omega_\mathrm{seed}$, 
which seeds further non-degenerate FWM-based cascaded processes with non-harmonic frequencies:
$\omega_\mathrm{SWM}^\mathrm{as} = \omega_\mathrm{FWM}^\mathrm{as}+ \omega_\mathrm{pump} -\omega_\mathrm{seed} =3\omega_\mathrm{pump} -2\omega_\mathrm{seed}$ and
$\omega_\mathrm{EWM}^\mathrm{as} = \omega_\mathrm{SWM}^\mathrm{as}+ \omega_\mathrm{pump} -\omega_\mathrm{seed} = 4\omega_\mathrm{pump} -3 \omega_\mathrm{seed}$.
Here, the idlers on the anti-Stokes side are much stronger than those on the Stokes side, 
while the relative magnitude relationship is reversed in the conjugated case, $2\omega_\mathrm{pump} > \omega_\mathrm{seed} > \omega_\mathrm{pump}$.

The perturbative understanding described above, however, captures only part of the observed phenomena.
In fact, a DE in SCG in H$_2$O was successfully demonstrated using a non-harmonic two-color field, 
comprising a pump (1036~nm, 800~nJ) and a seed (1300~nm, 100~nJ) [Fig.~\ref{fig1}(a)].
The above perturbative picture predicts cascaded FWM processes with discrete spectral components at $\lambda_\mathrm{FWM}^\mathrm{as} = 861$~nm, 
$\lambda_\mathrm{SWM}^\mathrm{as} = 737$~nm, and $\lambda_\mathrm{EWM}^\mathrm{as} = 644$~nm.
Hereafter, we refer to the spectral range $\lambda_i^\mathrm{as} \pm 40$~nm as the ``$i$ region''.
Under the optimized condition used in Fig.~\ref{fig1}(a), however, a broadband enhancement about three orders of magnitude was observed in the 600--800~nm range, 
compared to the optimized single-color case, despite the latter employing a pump pulse with 1.5 times higher energy (1200~nJ, $1P_\mathrm{cr}$). 
Notably, although the seed pulse energy was only 100~nJ, the seed played a dramatic role in enabling broadband SCG, 
as evidenced by comparison with the spectrum obtained using the pump pulse alone at the same energy (800~nJ) [Fig.~\ref{fig1}(a)].

\begin{figure}[t]
\centering
\includegraphics[width=\linewidth]{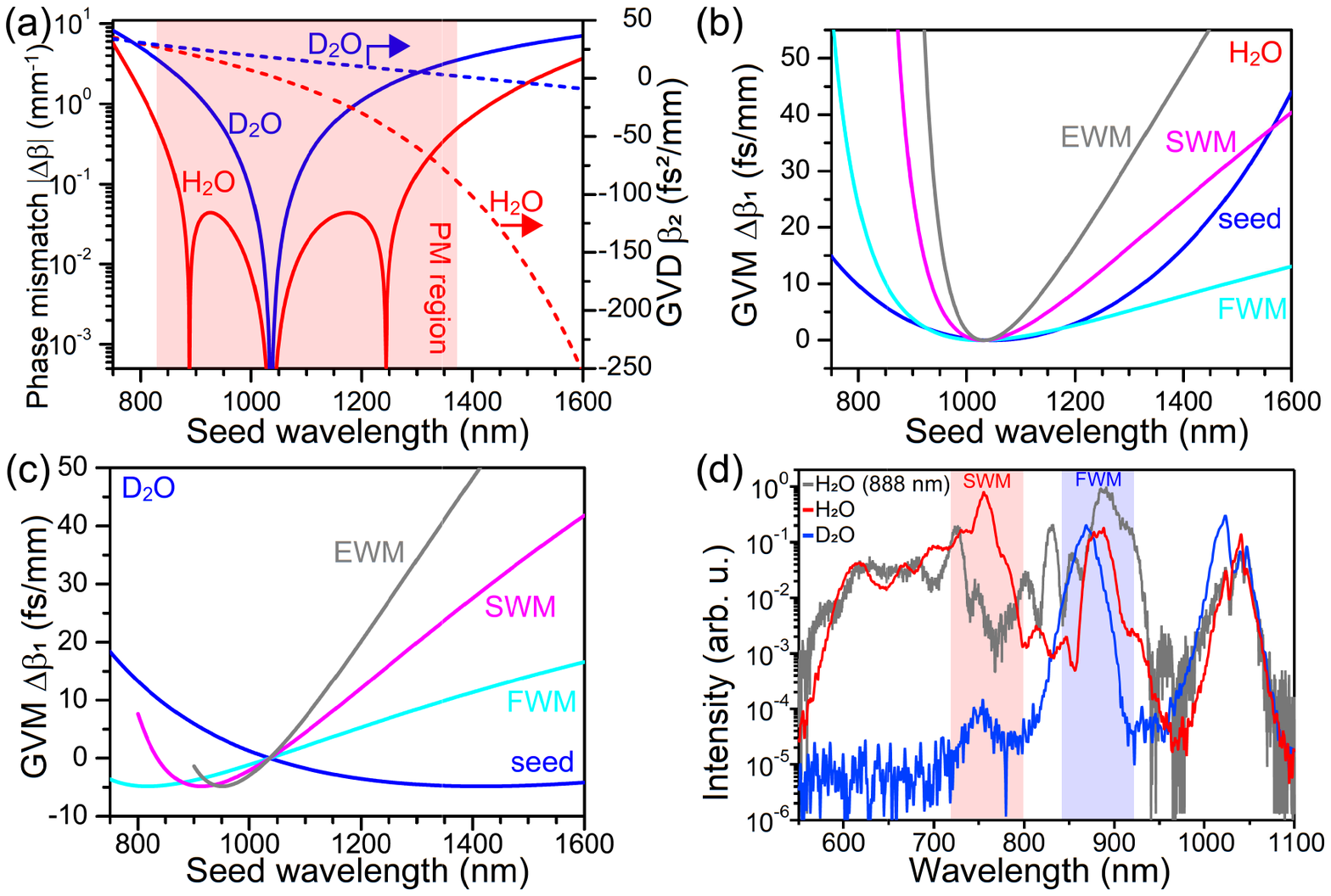}
\caption{{\small 
(a)  The absolute value of the phase mismatch $|\Delta \beta|$ for degenerate FWM ($\omega_\mathrm{FWM} = 2\omega_\mathrm{pump} - \omega_\mathrm{seed}$) in H$_2$O (solid red) and D$_2$O (solid blue), along with the GVD $\beta_2$ of H$_2$O (dotted red) and D$_2$O (dotted blue) \cite{Kedenburg2012}, 
plotted as functions of  $\lambda_\mathrm{seed}$. 
The PM condition in H$_2$O is satisfied at $\lambda_\mathrm{seed} = 1243$~nm and 888~nm. 
The ZDW of H$_2$O and D$_2$O are 1048~nm \cite{Dharmadhikari2016} and 1415~nm, respectively. 
The wavelengths of maximum group velocity are slightly offset from the ZDWs, appearing at 1051~nm for H$_2$O and 1411~nm for D$_2$O, due to the finite second derivative $d^2n/d\lambda^2$.
(b, c) GVM between the pump and the FWM (cyan), SWM (magenta), EWM (gray), 
and seed  (blue) pulses, 
plotted as functions of $\lambda_\mathrm{seed}$ in H$_2$O (b) and D$_2$O (c). 
Group velocity matching between the FWM and seed pulses is satisfied at $\lambda_\mathrm{seed} = 1176$~nm and 926~nm for H$_2$O.
(d) SC spectra generated in H$_2$O (red) and D$_2$O (blue) using $\omega_\mathrm{pump}$ (1036~nm, 800~nJ) and $\omega_\mathrm{seed}$ (1300~nm, 100~nJ). 
For comparison, the SC spectrum generated using the same pump and a conjugate seed (888~nm, 100~nJ), which satisfies the PM condition, is also shown (gray).
}}
\label{fig2}
\end{figure}

The key features of the underlying physics can be elucidated  by examining the dependence of the SC spectrum 
on the seed pulse energy $E_{\text{seed}}$, while keeping the pump energy fixed at 800~nJ.
As shown in Fig.~\ref{fig1}(b), even very weak seed pulses initiated nonlinear effects: 
FWM appeared 
at $E_{\text{seed}} = 1~\text{nJ}$, and SWM at 5~nJ, 
accompanied by an increase of more than two orders of magnitude in the FWM intensity. 
This nonlinear increase cannot be explained by perturbative FWM, 
which is a linear process with respect to the seed, as discussed later. 
The FWM signal began to saturate around 15~nJ, indicating that significant part of the pump energy was transferred to the SWM process via the cascaded FWM rather than direct SWM. 
Also around this seed energy (15 nJ), spectral broadening of the FWM and the SWM signal occurred. 
 Interestingly, this broadening showed a clear asymmetry toward shorter wavelengths, 
suggesting dominance of XPM induced by the falling edge of the pump pulse 
with larger group velocity than FWM and SWM [Fig.~\ref{fig2}(b)], 
rather than self-phase modulation (SPM), which typically results in symmetric broadening.
Figure~\ref{fig1}(c) shows the SC spectrum as a function of $E_{\text{seed}}$ up to 100~nJ, 
the stability limit for SCG.
The broadest spectrum was obtained at around 100~nJ, as also shown in Fig.~\ref{fig1}(a), 
whereas intuitively expected broadening into broadband light via EWM and XPM was not observed.


The delay ($\tau$) dependence of SCG in H$_2$O is shown in Figs.~\ref{fig1}(d,e), 
where $\tau$ is defined as the pump-seed delay, with $\tau > 0$ indicating that the pump precedes the seed.
The zero delay ($\tau = 0$) is defined as the condition where the SC spectral cutoff reaches its maximum extent,
and the same pulse parameters as in Fig.~\ref{fig1}(a) were used.
As expected, clear seed amplification associated with FWM was observed for $|\tau| \lesssim 160$ fs [Fig.~\ref{fig1}(d)], 
accompanied by the synchronized emergence of FWM and SWM components [Fig.~\ref{fig1}(e)]. 
The FWM and SWM signals exhibit a red shift that increases with $\tau$, 
an effect attributed to the positive chirp of the pump induced by SPM [Fig.~\ref{fig1}(e)].
In contrast, the center frequency of the seed was blue-shifted for $\tau > 0$ 
relative to that at $\tau = 253$~fs (virtually no temporal overlap), whose red spectral component was already 
absorbed by the second overtone or combination band of the OH stretching vibration 
in H$_2$O centered at 1450 nm. 
For $\tau < 0$, the seed was red-shifted owing to its negative chirp 
induced by H$_2$O, combined with amplification via FWM and soliton compression, 
as evidenced by the spectral broadening observed at $|\tau| \lesssim 53$~fs [Fig.~\ref{fig1}(d)].
 The seed gain of 2.5 was measured at $\lambda_{\text{seed}} = 1300~\text{nm}$, 
with no significant dependence on wavelength [Fig.~\ref{fig1}(f)]. 
This suggests that the amplification primarily results from degenerate FWM, rather than from SRS involving, for example, the bending mode of H$_2$O at $1650~\text{cm}^{-1}$, 
which can be resonantly excited at $\lambda_{\text{seed}} \approx 1250~\text{nm}$~\cite{Pattenaude2018}.

\begin{figure}[t]
\centering
\includegraphics[width=\linewidth]{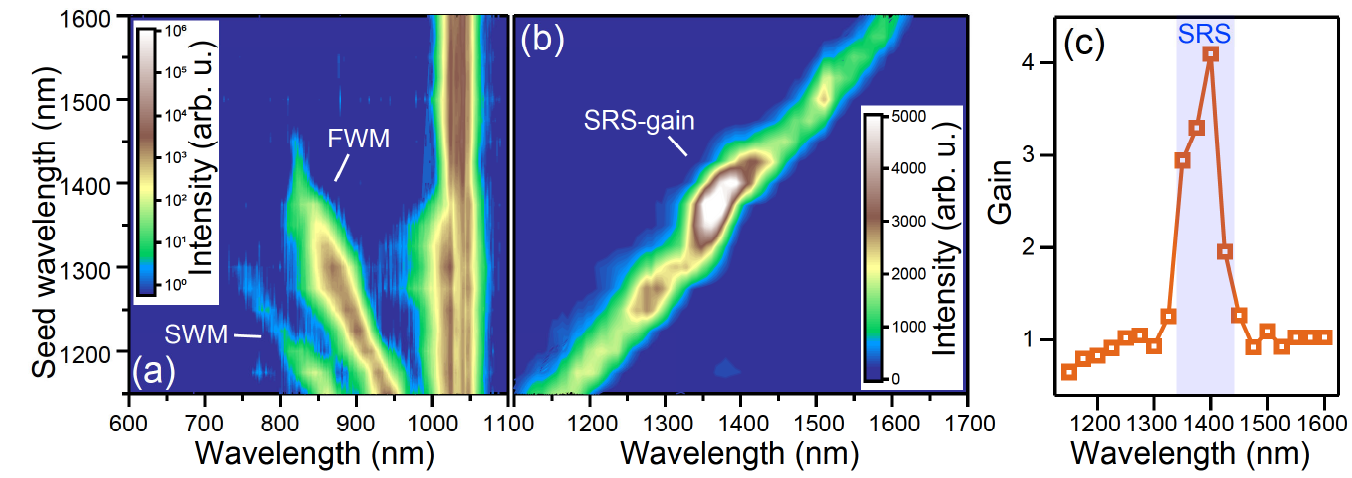}
\caption{{\small 
(a, b) SC spectra generated in D$_2$O as functions of $\lambda_\mathrm{seed}$, measured using (a) a Si-based and (b) an InGaAs-based spectrometer.  
(c) Total gain of the seed pulse in D$_2$O, plotted as a function of $\lambda_\mathrm{seed}$.  
In all measurements, the pump and seed pulse energies are fixed at $E_\mathrm{pump} = 800$~nJ and $E_\mathrm{seed} = 100$~nJ, respectively.
}}
\label{fig3}
\end{figure}


The observed dominance of FWM over SRS in H$_2$O in Fig.~\ref{fig1}(f) can be understood in terms of phase-matching (PM) of FWM \cite{Chen1990,Junaid2024}.
Figure~\ref{fig2}(a) presents the absolute value of the phase mismatch $|\Delta \beta|$ for degenerate FWM, alongside the group velocity dispersion (GVD) $\beta_2$ of H$_2$O and D$_2$O, 
plotted as functions of the seed wavelength. 
Notably, the PM condition in H$_2$O is satisfied at $\lambda_\mathrm{seed} = 888$ and 1243~nm, 
the latter close to the SRS resonance near $\lambda_\mathrm{seed} = 1250$~nm, effectively dominates SRS gain. 
Here, we refer to the region with $|\Delta \beta| \lesssim 0.5~\mathrm{mm}^{-1}$ for H$_2$O as the PM region 
($830~\mathrm{nm} < \lambda_\mathrm{seed} < 1375~\mathrm{nm}$).

The effect of PM of FWM is, in the present case, significantly enhanced by the group velocity matching. 
Figures~\ref{fig2}(b,c) show the group velocity mismatch (GVM) between pulse 
$i$ ($i = \mathrm{FWM}, \mathrm{SWM}$, etc.) and the pump, 
$\Delta \beta_1^{i,\mathrm{pump}} := 1/v_g^i - 1/v_g^\mathrm{pump}$, 
in H$_2$O (b) and D$_2$O (c) as functions of $\lambda_\mathrm{seed}$.
Here, $v_g^j$ denotes the group velocity of pulse $j$. 
In H$_2$O, group velocity matching between the seed and FWM signals is satisfied at $\lambda_\mathrm{seed} = 1176$~nm and 926~nm.
$\Delta \beta_1^{\mathrm{seed},\mathrm{pump}}$ and $\Delta \beta_1^{\mathrm{FWM},\mathrm{pump}}$ in the PM region remains sufficiently small ($<15$~fs/mm), 
relative to the pump pulse duration of 180~fs and the observed maximum filamentation length ($\approx 2$~mm), 
thereby enhancing the parametric FWM process in H$_2$O.
For D$_2$O, on the other hand, phase mismatch is significantly larger than that in H$_2$O [Fig.~\ref{fig2}(a)], 
and no spectral enhancement was observed, as shown in Fig.~\ref{fig2}(d). 
This observation is counterintuitive, when focusing only on its higher nonlinear refractive index $n_2$ ($6.4 \times 10^{-16}~\mathrm{cm}^2/\mathrm{W}$~\cite{Tcypkin2021}) 
compared to that of H$_2$O ($1.9 \times 10^{-16}~\mathrm{cm}^2/\mathrm{W}$~\cite{Tcypkin2021} or 
$4.1 \times 10^{-16}~\mathrm{cm}^2/\mathrm{W}$~\cite{Boyd2019}).


In addition to PM and GVM, soliton dynamics plays crucial role in this DE process.
The zero-dispersion wavelengths (ZDWs) of H$_2$O (1048~nm) indicates that 
the seed wavelength at 1300~nm lies in the anomalous dispersion regime of H$_2$O, 
suggesting the possibility of soliton compression during the amplification via FWM accompanied 
by RDW emission~\cite{Leo2014,Genty2004,Vasa2014,Suminas2016,Panagiotopoulos2015,Heinzerling2025}. 
To logically isolate this effect, we reversed the roles of the seed and idler waves in FWM in Fig.~\ref{fig2}(d) by employing a seed at 888 nm, which also satisfies 
the PM condition to efficiently generate a FWM idler around 1243 nm. 
The gray curve in Fig.~\ref{fig2}(d) shows the SC spectrum obtained with this conjugate wavelength condition.
Despite the stronger amplified seed in the FWM region observed in this case, which is advantageous 
for generating the SWM component through cascaded FWM, 
a marked suppression in the SWM region is evident [Fig.~\ref{fig2}(d)]. 
In contrast, a DE is observed in the 600--730~nm range, similar to the results before applying conjugation.
This indicates that the spectral component in this range does not originate from the cascaded FWM process, 
but rather suggests the presence of another mechanism.

The two fundamental parameters of RDW emission strongly support that RDW emission 
is the responsible mechanism.
In fact, the PM condition for RDW emission,
$\beta(\omega_{\mathrm{RDW}}) = \beta(\omega_\mathrm{seed}) 
+ (\omega_{\mathrm{RDW}} - \omega_\mathrm{seed})/v_{g}^\mathrm{seed}$,
yields $\lambda_\mathrm{DW} = 666$~nm, which corresponds to the center of the observed 600--730~nm range. 
The soliton order, on the other hand,
$N = \sqrt{\gamma P_0 T_0^2 / |\beta_2|}$,
was estimated to be $N \approx 1.2$, 
using $T_0 = 36$~fs (characteristic pulse duration at 1300~nm), 
$P_0 \approx 2.4$~MW (peak power of a 2.5-times-amplified 100~nJ seed after 39\% absorption in H$_2$O), 
and $\gamma = 2\pi n_2 / (\lambda_0 A_\mathrm{eff})$ 
with $A_\mathrm{eff} = \pi w_0^2$.
This value satisfies the condition for RDW emission, and indeed, significant spectral broadening of the seed is observed for $|\tau| \lesssim 53$~fs in Fig.~\ref{fig1}(d) 
accompanied by the enhanced signal observed 600--730~nm range in Fig.~\ref{fig1}(e).
Here, the FWM process $2\omega_\mathrm{FWM}-\omega_\mathrm{seed}$, 
which could also generate spectral component in this region, is expected to make only a minor contribution 
 due to the large phase mismatch ($\approx 36$ mm$^{-1}$), 
which is likely the origin of the spectral modulation observed in this range. 
For pulse-duration measurements of SC, however, phase stabilization of each pulse is required, 
and modifications to the experimental setup are currently in progress.


Finally, we return to the issue of competition between FWM and SRS. 
Figure~\ref{fig3}(a,b) shows the SCG spectra generated in D$_2$O as functions of $\lambda_\mathrm{seed}$. 
The pulse energies of the seed and pump are identical to those used in Fig.~\ref{fig1}(a). 
A clear enhancement is observed at $\lambda_\mathrm{seed} = 1350$--1425~nm, which corresponds to resonance with the stretching mode of D$_2$O (2300--2700~cm$^{-1}$). 
The gain reaches 4.1 at 1400~nm [Fig.~\ref{fig3}(c)]; however, no enhancement is observed for FWM or SWM. 
In contrast to the H$_2$O case [Fig.~\ref{fig1}(f)], SRS dominates over FWM in D$_2$O 
because of the large phase-mismatch for FWM and the high gain coefficient of SRS \cite{Chen1990,Junaid2024}.


In summary, we have demonstrated significantly enhanced SCG in H$_2$O using non-harmonic two-color laser fields.
Compared to single-color or harmonic two-color configurations, these non-harmonic fields enable much broader spectral generation with substantially higher efficiency. 
This enhancement arises from the combined effects of soliton compression accompanied by RDW emission, 
phase- and group-velocity-matched cascaded FWM, and subsequent XPM. 
This work not only  establishes a robust approach to achieving broader and more intense SCG, but also
provide new potential for broadband/multicolor spectroscopy
in liquids--particularly in H2O--a rapidly emerging 
frontier in attosecond science and surface/interface physics.

\begin{backmatter}
\bmsection{Funding} 
This study was supported by JSPS KAKENHI Grant-in-Aid for
Scientific Research (A) [22H00296],
 Grant-in-Aid for Scientific Research (B) [23H01877 and 23K26570],
Grant-in-Aid for Challenging Research (Exploratory) 
[21K18896], 
ATLA Japan Innovative Science and Technology Initiative for
Security [JPJ004596], JST-CREST 
[JPMJCR22L2], MATSUO FOUNDATION, Amada Foundation General research and development grant
[AF-2021212-B2, AF-2022234-B3], 
the International Research Exchange Support Program of the National Institutes of Natural Sciences (NINS),
Joint Research by NINS [01112104], and
by the Special Project by Institute for Molecular Science [IMS programme 22IMS1101].

\bmsection{Disclosures} The authors declare no conflicts of interest.

\vspace{-2mm}

\end{backmatter}

\bibliography{kanaiOL20250606}



\end{document}